\documentstyle[12pt,bezier]{article}
\newfont{\mathea}{msam10 scaled\magstep0}
\newfont{\matheb}{msbm10 scaled 1095}
\newfont{\tmpEins}{cmsy10 scaled 2074}
\newfont{\tmpZwei}{cmsy10 scaled 1095}
\newfont{\tmpDrei}{cmsy10 scaled 1000}
\newfont{\tmpVier}{cmsy5 scaled 1000}
\newfont{\tmpFuenf}{msbm7 scaled\magstep0}

\def\Bbb#1{\mathchoice{\mbox{\matheb #1}}{\mbox{\matheb #1}}%
 {\mbox{\tmpFuenf #1}}{\mbox{\tmpFuenf #1}}}

\def\dach#1#2{\mbox{$\mathop{\vbox{\ialign{%
  $##\crcr\hfil #1 \hfil$\crcr}}}\limits^{\scriptscriptstyle #2}$}}

\def\rnzs{\dach{\rho_2}{\mbox{$\scriptscriptstyle\kern-.7mm0$}}\kern-1.2mm'}

\def\Subset{\mbox{$\subset\kern-.5mm\subset$}}
\newcommand{\LI}{\mbox{{\rm L$^{\kern-.15em\raise.2ex\hbox{\scriptsize 1}}$}}}

\def\Ldummy{\left.\bgroup}
\def\Rdummy{\egroup^{\rule{0mm}{1.4mm}}\right.}
\def\LA{\left\langle\bgroup}
\def\RA{\egroup^{\rule{0mm}{1.4mm}}\right\rangle_{\cal A}^{}}
\def\LR{\left(\bgroup}
\def\RR{\egroup^{\rule{0mm}{1.4mm}}\right)}
\def\LG{\left\{\bgroup}
\def\RG{\egroup^{\rule{0mm}{1.4mm}}\right\}}

\def\Wort#1{\mbox{{\rm #1\kern.1em}}}

\def\lfac#1#2{\vcenter{\hbox{$#1\kern-.2em\raise-.6ex\hbox{\Large{/}}%
 \kern-.2em\raise-1.2ex\hbox{$#2$}$}}}

\def\gin{\mbox{\tmpZwei\symbol{91}\kern-1.4mm\rule{.2mm}{1.85mm}\kern1.4mm}}
\def\gni{\mbox{\tmpZwei\symbol{92}\kern-1.4mm\rule[.15mm]{.2mm}{1.85mm}%
  \kern1.4mm}}

\def\EINS{{\mathop{1\kern-.25em\mbox{{\rm{\small l}}}}}}

\begin{document}

\Large Remarks related to the paper of Rafael de la Madrid:
 "On the inconsistency of the Bohm-Gadella theory
with quantum mechanics", JPhysicsA 39, No. 29, 9255-9268 (2006)

\vspace{1cm}

\normalsize Hellmut Baumg\"artel

\vspace{3mm}

Mathematical Institute, University of Potsdam

Am Neuen Palais 10, PF 601553

D-14415 Potsdam, Germany

e-mail: baumg@rz.uni-potsdam.de

\vspace{1cm}

\begin{abstract}
The paper contains critical comments to the paper mentioned in the title from the 
mathematical point of view.
\end{abstract}

\vspace{1mm}

The following remarks refer to the paper mentioned in the title. 
In the following it is quoted as [R].
They concern the pure mathematical point of view.
There are critical comments to special points (3,4,5) of that paper and
to statements which could cause misunderstandings. The paper contains even bad mistakes.
This special critique leads to the conclusion that the paper fails its own aim
in the following sense: The author of these remarks agrees completely with the
following statements of Rafael de la Madrid in the introduction of his paper,

"... the resonance states and time asymmetry can be achieved within standard
quantum mechanics."

"...the content of the Hardy axiom is not a matter of assumption, but 
a matter of proof."

Now first the comment to point 3 shows that the "Hardy axiom" can be proved rigorously 
within the framework of standard quantum mechanics, i.e. well-understood
it is not an axiom but a fact, a theorem. Second, the comments to point 4 suggest that,
mathematically speaking, "time asymmetry" is an intrinsic element of the
mathematical apparatus of standard quantum mechanics which is finally due
to the semiboundedness of the Hamiltonians and the property that {\em their
absolutely continuous spectrum is of homogeneous multiplicity} and (in general)
coincides with the positive half line. Independently of the special shape of what 
Rafael de la Madrid calls "Bohm-Gadella theory" and of special objections one can have
against it (cf. for example the remark in this letter
concerning the extensive use of the
Lippman-Schwinger equation in this connection) it seems to be a merit of
Bohm and Gadella to have perceived that the Hardy spaces are decisive for these
connections, which obviously do not leave the framework of standard quantum mechanics.

\vspace{2mm}

To begin with the context
let $H$ be the selfadjoint operator on the Hilbert space ${\cal H}_{+}:=
L^{2}(0,\infty)$ given by the differential expression
\[
(Hf)(r):=-\frac{d^{2}f}{dr^{2}}(r) + V(r)f(r),\quad f\in{\cal H}_{+},
\]
together with the boundary condition $f(0)=0$, where $V$ is real-valued, locally integrable,
$V(r)=0$ for $r>R>0$ and 
$\int_{0}^{R}r\vert V(r)\vert dr<\infty.$ If $V(r)\geq 0$ then $H$ has no eigenvalues.
The case in [R] is a special case of this setting. 
Further let $H_{0}$ be the selfadjoint operator
of the same type but with $V=0$. The wave operators
\[
W_{\pm}:=\mbox{s-lim}_{t\rightarrow\pm\infty}e^{itH}e^{-itH_{0}}
\]
exist and  they are
asymptotically complete. Let $r\rightarrow \phi(r,E)$ be the so-called regular 
solution of the differential equation
\[
-\frac{d^{2}y}{dr^{2}}(r)+V(r)y(r)=Ey(r),\quad \phi(0,E)=0,\;\phi'(0,E)=1,
\]
and $\phi_{0}(\cdot,E)$ the corresponding solution for $V=0.\,\phi$ is an entire
function in $E$, for example $\phi_{0}(r,E)=\frac{\sin\sqrt{E}r}{\sqrt{E}}.$
For the calculation of the
corresponding unitary canonical spectral representations of $H$ and $H_{0}$ 
one has to use the so-called Jost functions $A_{\pm}(\cdot)$, given by
\[
\phi(r,E)=A_{-}(E)e^{i\sqrt{E}r}+A_{+}(E)e^{-i\sqrt{E}r},\quad r>R,\,E>0.
\]
For example, for $\phi_{0}$ one has $A_{-}(E)=\frac{1}{2i\sqrt{E}},\,
A_{+}(E)=-\frac{1}{2i\sqrt{E}}$. Then
\[
{\cal H}_{+}\ni f\rightarrow\Psi f\in{\cal H}_{+}:
(\Psi f)(E):=\frac{1}{2\sqrt{\pi}E^{1/4}\vert A_{+}(E)\vert}\int_{0}^{\infty}\phi(r,E)f(r)dr
\]
and
\[
(\Psi_{0}f)(E):=\frac{E^{1/4}}{\sqrt{\pi}}\int_{0}^{\infty}\phi_{0}(r,E)f(r)dr,
\]
such that
\[
\Psi(e^{-itH}f)(E)=e^{-itE}(\Psi f)(E),\quad \Psi_{0}(e^{-itH_{0}}f)(E)=
e^{-itE}(\Psi_{0}f)(E).
\]
For convience we denote the multiplication operator $g(E)\rightarrow Eg(E),\,g\in
{\cal H}_{0}$ by $M$. For the (unitary) inverse transformations $\Psi^{-1},\,
\Psi_{0}^{-1}$ one obtains
\begin{equation}
(\Psi^{-1}g)(r)=\frac{1}{2\sqrt{\pi}}\int_{0}^{\infty}\phi(r,E)g(E)
\frac{1}{E^{1/4}\vert A_{+}(E)\vert}dE,\quad g\in{\cal H}_{+},
\end{equation}
\[
(\Psi_{0}^{-1}g)(r)=\frac{1}{\sqrt{\pi}}\int_{0}^{\infty}\frac{\sin(\sqrt{E}r)}
{E^{1/4}}g(E)dE,\quad 
g\in{\cal H}_{+}.
\]
W.r.t. the canonical spectral representations the wave operators $W_{\pm}$ act by
wave matrices $E\rightarrow W_{\pm}(E)$ as multiplication operators, defined by
$\Psi W_{\pm}\Psi_{0}^{-1}.$

The so-called Lippman-Schwinger equation yields an equation for the wave matrices which
is of limited value for concrete calculations, in general. (Also for this reason it is
not recommendable to use this equation as a starting point for proposals of
new postulates with a deeper conceptional aspect.) Using the formula
\[
W_{\pm}f=\pm i\,\mbox{s-lim}_{\epsilon\rightarrow+0}
\int_{0}^{\infty}E(d\lambda)(\epsilon R_{0}(\lambda\pm i\epsilon))f,
\]
for the wave operators
where $E(\cdot)$ denotes the spectral measure of $H$ and $R_{0}(z):=(z-H_{0})^{-1}$ 
the resolvent of $H_{0}$, one obtains for the wave matrices
\[
W_{\pm}(E)=\mp i\frac{A_{\pm}(E)}{\vert A_{+}(E)\vert},\quad E>0,
\]
i.e. the wave matrices coincide (up to a normalization factor) with the Jost functions. Considering
${\cal H}_{+}$ as the space of the spectral representation of $H$ and putting
\[
\Phi_{\pm}:=P_{+}{\cal H}^{2}_{\pm}\subset {\cal H}_{+},
\]
where ${\cal H}^{2}_{\pm}\subset{\cal H}:=L^{2}(\Bbb{R})$ denote the Hardy spaces w.r.t.
the upper resp. lower half plane and $P_{+}$ the projection by multiplication
with the characteristic function $\chi_{[0,\infty)}$ such that 
${\cal H}_{+}=P_{+}{\cal H}$, then it turns out that $\Phi_{\pm}$ is a dense linear
manifold in ${\cal H}_{+}$ and $\tilde{\Phi_{\pm}}:=\Psi^{-1}\Phi_{\pm}$ is dense in
${\cal H}_{+}$, i.e. each "radial function" $f_{\pm}\in\tilde{\Phi_{\pm}}$ gives
via $\Psi f_{\pm}$ the "positive part" of a Hardy function in ${\cal H}_{+}$,
considered as the space of the spectral representation of $H$. Conversely, if the positive
part of a Hardy function is given, the corresponding "radial function" can be calculated
by (1). This is a comment to point 3 of [R]. It shows that - in contradiction to
the assertion of Rafael de la Madrid that "the limits (2.18) and (2.19) 
are in general not zero" - there
is a dense set of radial functions from $L^{2}(0,\infty)$ (wave functions) which
produce a (dense) set of positive parts of 
Hardy functions in the spectral representation space (again
$L^{2}(0,\infty)$), in particular a dense set which are additionally Schwartz
functions. Therefore, the conclusion of "inconsistency" in [R] is nonsense.

\vspace{3mm}

The dense manifolds $\Phi_{\pm}$ are not invariant w.r.t. $e^{-itM}$, in general. 
However $\Phi_{+}$ is invariant for $t\leq 0$ and $\Phi_{-}$ is invariant for
$t\geq 0$.

The multiplication operator $M: Mf(E)=Ef(E)$ can be extended to the 
whole space ${\cal H}$ such that
\[
{\cal H}\ni g\rightarrow e^{-itM}g: e^{-itM}g(E)=e^{-itE}g(E).
\]
Note that for the extended (spectral) evolution the {\em subspaces} ${\cal H}_{\pm}^{2}$
are invariant for $t<0$ resp. $t>0$. Now $e^{-itM}$ is the Fourier transform
of the shift transformation $T(t)$ on ${\cal H},\,T(t)g(x):=g(x-t)$, i.e.
\[
F^{-1}e^{-itM}F=T(t),
\]
where the Fourier transformation is given by
\[
Fg(E):=\frac{1}{\sqrt{2\pi}}\int_{-\infty}^{\infty}e^{-iEx}g(x)dx.
\]
Note that
\[
F^{-1}Q_{\pm}F=P_{\mp},
\]
where $Q_{\pm}$ is the projection onto the Hardy space ${\cal H}^{2}_{\pm}$ and $P_{-}$
the projection by multiplication with the characteristic function
$\chi_{(-\infty,0]}(\cdot).\,P_{\mp}{\cal H}$ are the wellknown incoming/outgoing
subspaces of the shift evolution $T(t)$. The connection between the evolution $e^{-itH}$
and $T(t)$ is then given by
\[
e^{-itH}=\Psi^{-1}P_{+}FT(t)F^{-1}P_{+}\Psi
\]
and the invariant manifolds for $t>0,\,t<0$ are
\[
\Psi^{-1}\Phi_{\mp}=\Psi^{-1}P_{+}{\cal H}^{2}_{\mp}=\Psi^{-1}P_{+}FP_{\pm}{\cal H},
\]
i.e. $f_{\pm}\in\Psi^{-1}\Phi_{\mp}$ is given by
$f_{\pm}=\Psi^{-1}P_{+}Fg_{\pm}$ where $g_{\pm}\in P_{\pm}{\cal H}$, that is
$g_{\pm}$ are outgoing/incoming vectors w.r.t. the shift evolution. The correspondence
$f_{\pm}\leftrightarrow g_{\pm}$ is a bijection so that $g_{\pm}$
can be considered as a {\em representer} of $f_{\pm}$. In other words,
$g_{\pm}=F^{-1}\Psi(f_{\pm})$ or
\[
g_{\pm}(x)=\frac{1}{\sqrt{2\pi}}\int_{-\infty}^{\infty}e^{ixE}(\Psi f_{\pm})(E)dE
\]
This comments point 4 in [R]. The functions in equations (4.5), (4.8) are the
{\em representers} $g_{\mp}$ of the elements $f_{\mp}$ which are from the invariant
manifolds w.r.t. the "real" evolution $e^{-itH}.$ Obviously in [R] there
is a confusion between the parameter $t$ of the evolution and the variable in the
argument of the functions where the shift evolution acts (e.g. in (4.5)) which is
denoted by $x$ in the comment. (Already the "conclusion" equation (4.9) should
suggest that something is wrong in the starting statement.)

\vspace{3mm}

The statement in point 5: "... Hardy functions are not suitable for systems
whose spectrum is bounded from below" is definitely wrong. On the contrary,
they seem to be decisive for the connection of the "Gamov vectors", which are
special eigenvectors of the so-called "decay semigroup" of the Toeplitz type,
to the eigenlinear forms of
the resonances if their spectral theoretical characterization  w.r.t. the quantum
mechanical evolution is established. For example, in the case of the
finite-dimensional Friedrichs model on the positive half line this is pointed out in
[1] (see also [2]).

\vspace{1cm}

REFERENCES
\begin{enumerate}
\item H. Baumg\"artel: Rev. Math. Phys. 18, 61 - 78 (2006)\\
\item H. Baumg\"artel: The eigenvalue problem for resonances of the infinite-dimensional
Friedrichs model on the positive half line with Hilbert-Schmidt perturbations\\
arXiv: math-ph 0608036 (2006)
\end{enumerate}

\end{document}